# DFT study of structural, electronic and optical properties of 2D MgO monolayer under bi-axial mechanical strain

Kamal Kumar[1], Anjali Kumari[1], Soni Mishra[2], Ramesh Sharma[3], Abhishek Kumar Mishra[1]*

[1]Department of Physics, Applied Science Cluster, University of Petroleum and Energy Studies, Bidholi via Premnagar, Dehradun, Uttarakhand 248007, India

[2]Department of Physics, Graphic Era Hill University, Dehradun 248002, India

[3]Department of Applied Science, Feroze Gandhi Institute of Engineering and Technology, Raebareli, 229001, Uttar Pradesh, India

**Abstract**

The structural, electronic, and dielectric (optical) properties of graphene-like 2D MgO monolayer have been explored through first-principles calculations under bi-axial tensile and compressive mechanical strain within a range of -10% to +10%. Our findings revealed that the pristine MgO monolayer is an indirect band gap semiconducting material and the semiconducting mature of MgO monolayer remains consistent under both compressive and tensile mechanical strain. This nature of MgO is confirmed through partial density of states (PDOS) as well as electronic band structure. PDOS exhibits the contribution of different atomic orbitals in bond formation and nature of bond, while band structure provides insight into electron transitions between energy levels of valance and conduction bands. All optical parameters (dielectric function, reflectivity, energy loss, refractive index, extinction coefficient and absorption) are plotted in an energy range 0-15 eV. Within this energy interval, MgO possesses the highest value of the refractive index (2.13) at 3.12 eV energy. Also, a detailed analysis of changes in the geometrical structure of MgO monolayer is provided.

*Corresponding author: mishra_lu@hotmail.com, akmishra@ddn.upes.ac.in

**Introduction**

Reducing the dimensions of bulk (3D) materials to nano (nm) range (1nm-100 nm) is important for the advancement of materials used in various devices [1–3]. At this small range, the characteristics (electrical conductivity, melting points, chemical reactivity etc.) of materials alter abruptly[4–6]. This process of confining the dimensions of materials is called quantum confinement and the materials with these reduced dimensions are called nanomaterials[7–9]. The classification of nanomaterials into zero-dimensional (0D/quantum dots), one-dimensional (1D/nanowire) and two-dimensional (2D/nanosheets) materials depends on the no. of reduced dimensions [10–12]. All three dimensions of 0D materials (quantum dots) are in nanometer range, while 1D (nanotubes and nanowires) and 2D (thin films and nanosheets) nanomaterials have only two and one dimensions in nanometer range [13–15]. Due to high flexibility, immense mechanical strength, large surface to volume ratio and ultra-thin profile [16–18].2D materials are well-suited for diverse range of applications including energy storage, catalysis and electronics [19–21]. The study of remarkable characteristics of graphene has ignited keen curiosity and interest of scientific community to explore novel 2D nanomaterials like silicene, phosphorene, plumbene, sci-graphene, antimonene, arsenene, borophene and more [22–25]. Despite of spectrum of applications, these materials also face various limitations, that can be addressed by modifying their physical and chemical properties through different theoretical and experimental engineering techniques [26–28]. The selection of engineering technique for tailoring the properties of materials depends on time consumption, cost, scale of production, physical state of materials and applications [29,30]. These techniques include chemical functionalization, ion implantation, doping, strain engineering, defect engineering, laser irradiation and plasma treatment. Strain engineering is a versatile tool to effectively modify magnetic, optical, electronic and mechanical properties of materials under different strain conditions (in plane and out plane) [31–34]. The source of strain in material is not only external forces, but also the lattice mismatch between substrate and synthesized materials [35–38]. Hence, the understanding the effect of strain on optical and electronic properties of newly predicted materials is an important task [39–41]. The presence of strain changes the bond angle and interatomic distance within the atomic structure and this modification directly affects the distribution of energy levels in band structure leading to the change in electronic band gap [42–45]. On applying strain on materials, the interaction between lattice vibrations i.e., phonon and charge carriers change [46,47]. These changes in phonon behavior due to applied mechanical strain affects the charge carrier mobility [48–50]. Also, the changes induced by strain in a lattice lead to optical anisotropy due to which the optical properties of materials changes in different directions [51,52]. This anisotropy arises due to the change in symmetry of materials on applying strain [53,54]. A significant amount of literature is available examining the change in structural, optical and electronic properties of different materials in the presence of mechanical strain [43,44,47,55]. Kang et al. investigated the electrical and optical properties of pristine and strained graphyne from first principles calculations. Their study reveals that the band gap of graphyne decreases on applying compressive strain, while on applying tensile strain, band gap of graphyne increases. However, in both situations (tensile and compressive strain) the band gap of graphyne remains direct in nature [56]. Mohan et al. studied the dielectric and electronic properties of 2D silicene under uni and bi-axial strain. They observed a band gap opening up to 335 meV in silicene at small value of strain (4%), while a transition in metallic nature takes place when the applied strain exceeds 8%. Also, from the imaginary part of dielectric function they observed that under both uni and bi-axial tensile strain, interband-transitions are red shifted, while for compressive strain, these bands are bule shifted [57]. Wei et al. explored the optical properties and electronic structure of antimonene in the presence of biaxial strain. They observed a decreasing trend of static permittivity of antimonene under compressive strain. While static permittivity increases gradually under tensile strain. Furthermore, the speed of light in antimonene nanosheet slows down under compressive strain and light propagates faster under tensile strain [58].

2D Metal oxides (MO) are generally used in various technological applications, including microelectronics and catalysis [59–61]. For the systematic progress of this domains, it is necessary to have basic understanding of the characteristics of these materials [62]. This necessity has increased the interest in properties of (MOs) in last few decades [62]. Traditionally, MOs are catalytically inert promoters and often need some supporting materials (like $Al_2O_3$) to increase the surface area of an active materials [63,64]. These factors promoted various theoretical and experimental studies to explore the catalytic characteristics of real catalysts through single metal surface as a model [64]. MOs can be derived from variety of metals of periodic table including alkali metals, alkaline metals, precious metals, transition metals, each exhibits unique properties and applications [64,65]. Layered MOs are earning significant interest for different energy applications particularly in oxygen reduction and evolution process [66].

2D monolayer of alkaline earth MOs possess high reactivity, basic nature, high thermal stability and semiconducting nature due to which this class of MOs is useful in soil conditioning, water treatment and semiconducting industry [67–70]. In past few years, magnesium oxide (MgO) has explored widely in both experimental and theoretical research [71–73]. Different phases of bulk MgO are identified ($\beta_1$, $\beta_3$, $\beta_4$) and numerous efforts have been made to explore its structure of MgO thin films on metal substrates [74]. Zheng et al. theoretically predicted the graphene-like 2D monolayer of MgO (see **Figure 1**) and confirmed its semiconducting nature with an indirect band gap and high dynamical stability [75]. Subsequently, Menderes et al. predicted fascinating piezoelectric properties of 2D II and VI compounds [76]. Moghadam et al. studied the optical and magnetic properties of F, N, C and B doped MgO monolayer. They found that the doping of single N, C and F atoms generates magnetization in MgO monolayer which results magnetic and half-metallic behaviour, while doping of single F atom promotes non-magnetic characteristics[77]. Hoat et al. studied the effect of functionalization (N and F) on structural, optical and optical properties of MgO. They found that surface functionalization induces a buckling in planar MgO monolayer and allows MgO monolayer to adsorb light through wide range spam of wavelength (infrared to ultraviolet)[78].

Here in this study, we have studied impact of uniform biaxial strain on structural, electrical and optical characteristics of 2D MgO monolayer.

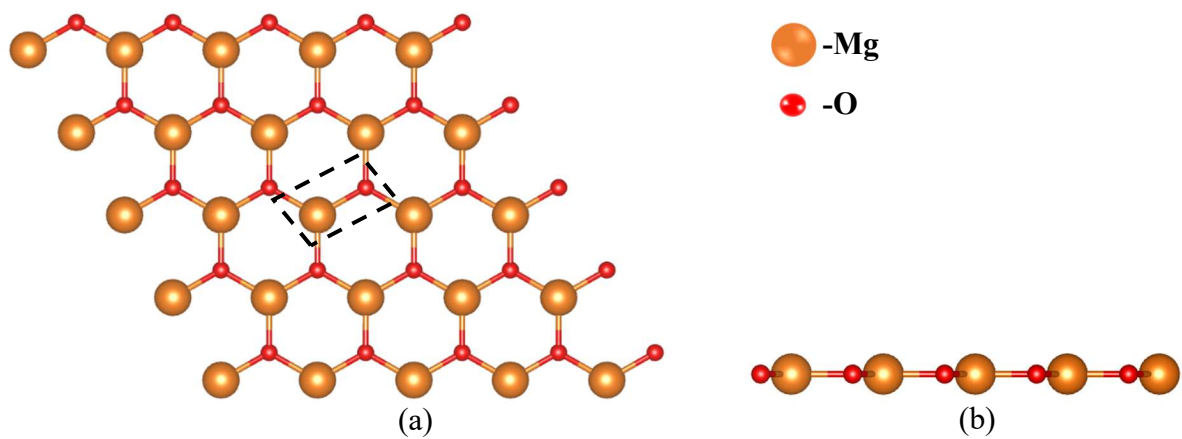

**Figure 1** Top view (a) and side view (b) of MgO monolayer (Orange and red balls represents Mg and O atoms respectively)

## Results and Discussion

### Structural Properties

The geometrical arrangement of 2D MgO monolayer is a graphene like flat hexagonal honeycomb structure formed by magnesium (Mg) and oxygen (O) atoms with P6m2 space group (no. 187)[78]. In figure 1, the atomic arrangement of 5×5×1 supercell of MgO monolayer is displayed. Our calculated lattice parameter after structural optimization is 3.29 Å, aligning good with earlier theoretical data[78]. The interatomic bond and bond angle formed between Mg and O atoms after relaxation are 1.90 Å and 120° respectively reflecting the planar nature of MgO monolayer. As discussed in introduction section, we have investigated effect of strain on structural properties of MgO monolayer. Hence, in **Table 1,** we have mentioned the optimized cell parameters along bond length under both positive and negative strain values. From **Table 1** it is clear that as the magnitude of applied negative strain increases, the bond length of Mg and O atoms decreases and at -10% strain value it becomes 1.71 Å, which is 0.2 Å less than the bond length without strain (at 0% strain). The reason behind this reduction is compression in MgO monolayer due to which Mg and O atoms are pushed together and bond length decreases. Conversely, the bond length continuously expands as the magnitude of applied positive strain increases. At +10% strain value, the bond length of MgO expands 5.79% of its original value (without strain). The cause of this expansion is that positive strain stretches MgO monolayer, pulling Mg and O atoms outwards, which results an increment in bond length.

**Table 1** Optimized cell parameters (a and b), band gap $E_g$, bond length (d) of MgO monolayer at different strain values

| Strain | a(Å) | b(Å) | d (Å) | $E_g$(eV) | Nature |
|---|---|---|---|---|---|
| **-10%** | 2.96 | 2.96 | 1.71 | 3.37 | Indirect |
| **-5%** | 3.12 | 3.12 | 1.80 | 3.28 | Indirect |
| **-2%** | 3.22 | 3.22 | 1.86 | 3.02 | Indirect |
| **0%** | 3.29 | 3.29 | 1.90 | 2.80 | Indirect |
| **+2%** | 3.35 | 3.35 | 1.94 | 2.69 | Indirect |
| **+5%** | 3.45 | 3.45 | 1.99 | 2.55 | Indirect |
| **+10%** | 3.62 | 3.62 | 2.01 | 2.25 | Indirect |

**Electronic Properties**

In this section, the electronic properties of bare and strained MgO monolayer are discussed through partial density of states (PDOS) (see **Figure 2**) and electronic band structure (refer to **Figure 3**). Both PDOS and band structure are drawn from -4eV and 7eV in which Fermi energy is fixed at zero. The absence of energy levels near Fermi level and a distinct separation between the energy bands in PDOS directly indicated the semiconducting nature of MgO monolayer. The lowest point of conduction band is found at Γ-point, while the highest point of valance band lies at K-point of the Brillouin zone. Therefore, no direct transition of electron is possible from valance to conduction band, confirming the indirect nature of band gap. Also, our band structure and PDOS profile exactly matches with the earlier reported pattern of Hoat et al. [78]MgO monolayer maintains this indirect and semiconducting nature of band gap in both strain conditions (positive and negative). As O atom exhibits higher electronegativity than Mg atom, Mg atom donates two electrons of its outer most s-orbital to p-orbital of O atom for completing its octate. That is why at each strain value, p-orbital of O atom give highest contribution in the upper part of valance band and s-orbital contributes in lower region of conduction band. As a result of this the chemical bond between Mg and O atoms in MgO monolayer is ionic in nature. On increasing the magnitude of applied positive strain, band gap of MgO decreases and at +10% biaxial strain it reduces to 2.25 eV which is 0.55 eV less than its value at 0% strain. This gradual reduction in electronic band gap of MgO monolayer can be observed in electronic band structure also. As the magnitude of applied biaxial tensile strain elevates, energy levels in the conduction band starts shifting towards Fermi level leading to separation gap between the energy bands. On the other hand, on increasing the magnitude of applied bi-axial compressive strain, the separation between the energy bands widens continuously due to levels of conduction band moving far away from Fermi energy. This movement of energy levels enhances the insulating characteristics of MgO monolayer. At -10% compressive strain, the gap between energy bands becomes 3.37 eV. These fluctuations in band gap of MgO are summarized in **Table 1**. The energy gap never disappears within the considered spam of mechanical strain values.

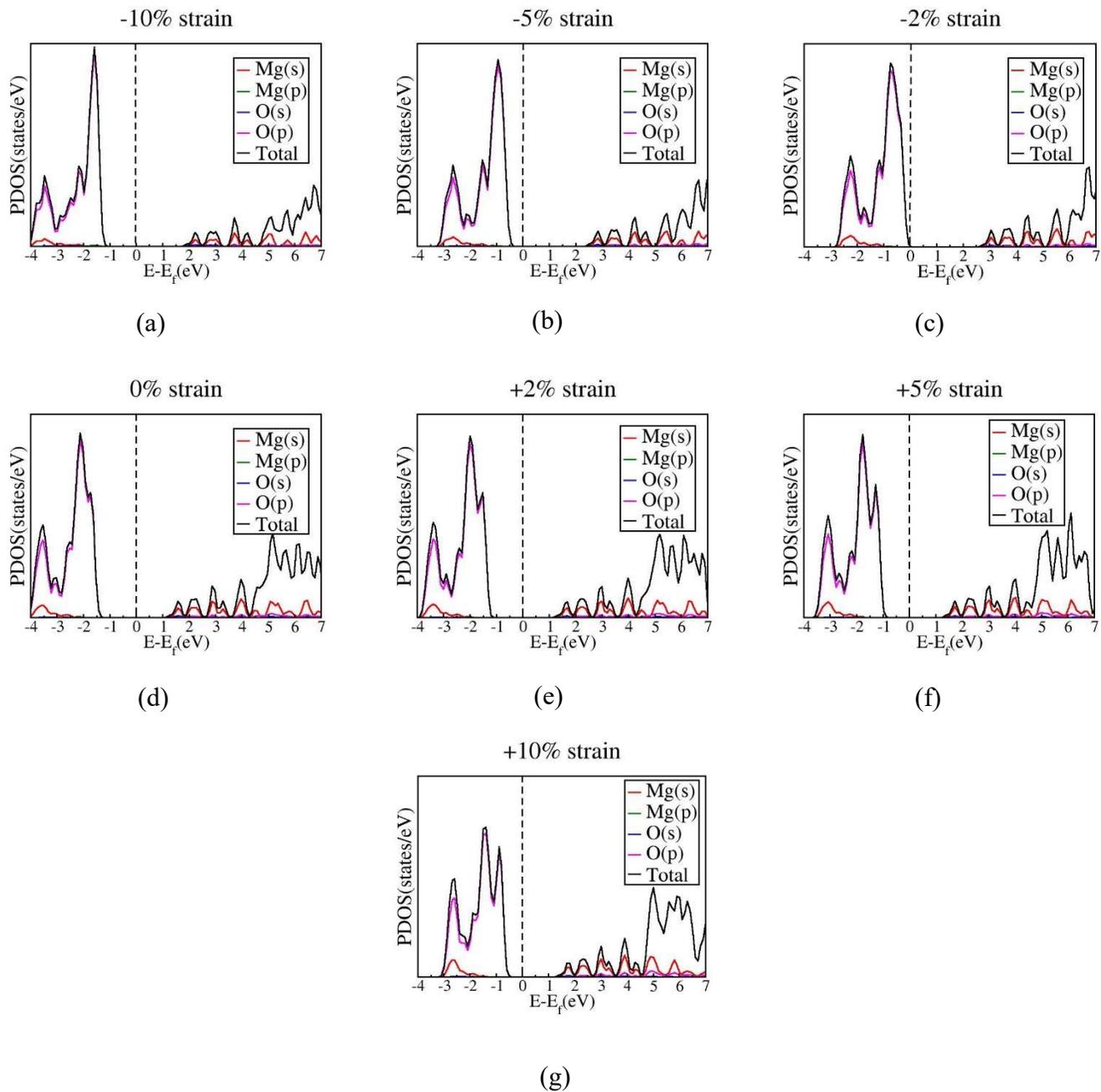

**Figure 2** PDOS of MgO monolayer under different strain values of applied biaxial mechanical strain

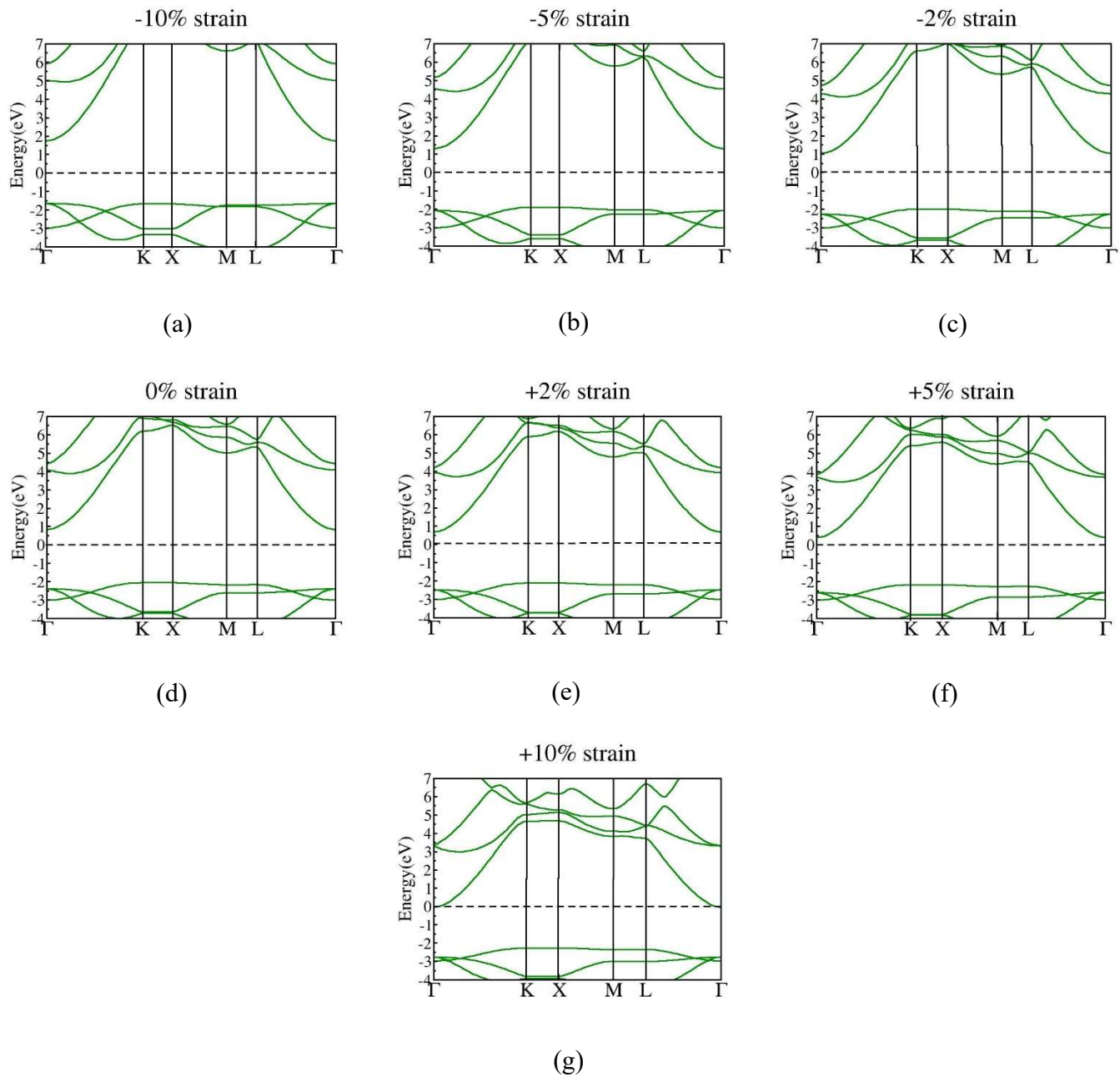

**Figure 3** Electronic band structure of pristine and strained MgO monolayer

## Optical Properties

In this section, we have discussed the optical properties of MgO monolayer to understand its interaction with electromagnetic radiation. As the real $\varepsilon_r(\omega)$ and imaginary parts $\varepsilon_i(\omega)$ of dielectric functions provides a comprehensive study of various optical properties of materials including absorption coefficient $\alpha(\omega)$, refractive index $n(\omega)$, extinction coefficient $k(\omega)$, average loss $L(\omega)$, and reflectivity $R(\omega)$, therefore, $\varepsilon_r(\omega)$ and $\varepsilon_i(\omega)$ are plotted in **Figure 4** at different biaxial strain values using PBE+GGA approximations. The computed optical parameters of the pristine MgO monolayer are summarized in **Table 2.** The most crucial factor for optical as well as dielectric properties of material is a complex dielectric function, which can be represented as equation (1)

$$\varepsilon = \varepsilon_r(\omega) + \hat{\imath}\ \varepsilon_i(\omega) \quad (1)$$

Where, $\varepsilon_r$ and $\varepsilon_i$ represent real and imaginary components of the dielectric function, respectively. Real component of dielectric function represents dispersion properties, polarization phenomenon and change in phase of light velocity when it travels through the material. Whereas, the imaginary component indicates the loss factor or absorption of light energy, indicating the amount of absorbed light of a specific frequency. The absorption phenomenon is linked with the transition of electrons within the energy levels. All dielectric properties are analyzed across an optical energy range from 0 to 15 eV depicted in figure 7(a) and 7(b).

The bandgap and static dielectric constant exhibit an inverse relationship through Penn's relation $\varepsilon_r(0) \approx 1 + (\hbar\omega_p/E_g)$. The Kramers–Kroning relation determines real part of the dielectric constant $(\varepsilon)$ by equation (2).

$$\varepsilon_r(\omega) = 1 + \frac{2}{\pi} P \int_0^\infty \frac{\omega' \varepsilon_2(\omega')}{\omega'^2 - \omega^2} d\omega' \quad (2)$$

Where, P is illustrating the principal integral, The imaginary part of dielectric function is give as

$$\varepsilon_i(\omega) = \frac{e^2 \hbar}{\pi m^2} \sum_{v,c} \int_{BZ}^\infty |M_{CV}(k)|^2\ \delta[\omega_{cv}(k) - \omega] d^3 k \quad (3)$$

We can calculate the reflectivity $R(\omega)$, and absorption coefficient $\alpha(\omega)$ by interpreting of $\varepsilon_r(\omega)$ and $\varepsilon_i(\omega)$. We can explore these relationships by employing the equations provided.

$$R(\omega) = \frac{[n-1]^2 + k^2}{[n+1]^2 + k^2} \quad (4)$$

$$n(\omega) = \sqrt{\frac{\{\varepsilon_r{}^2(\omega) + \varepsilon_i{}^2(\omega) + \varepsilon_r(\omega)\}}{2}} \tag{5}$$

We have computed imaginary and real parts of dielectric function against photon energy (eV) under biaxial mechanical strain spam of -10 % to +10 %. The highest values of real component ($\varepsilon_r$) of dielectric function under mechanical strain were observed to be 3.30(-10%), 3.37(-5%), 3.42 (-2%), 2.22(0%), 3.64 (+2%), 3.91(+5%), 4.36(+10%) at 8.66eV, 7.67 eV, 7.45eV, 3.91eV, 3.29eV, 3.15eV, 2.82eV energy values respectively for MgO monolayer. The maximum values of imaginary dielectric part ($\varepsilon_i$) under biaxial strain were observed 3.34(-10%), 3.84(-5%), 3.91 (-2%), 1.29(0%), 4.38 (+2%), 4.57(+5%), 4.33(+10%) at 8.81eV, 8.59 eV, 8.22eV, 8.21eV, 7.74eV, 7.48eV, 6.86eV respectively (see **Table 3**) for MgO monolayer as mentioned in the **Figure 4**. The maximum peak values of refractive index with the corresponding energies are 1.73(5.06eV), 1.77(3.79eV), 1.84(3.74eV), 1.48(3.92eV), 1.92(3.58eV), 2.03(3.43eV), 2.13(3.12eV) for MgO monolayer as shown in **Figure 5**. It is clear that, as we reduce the applied strain, refractive index decreases which allows the light to easily pass through the MgO monolayer. The static values of absorption coefficient with all applied strain were observed almost zero (0 %, 2 %, 5 %, 10 %, 2%, 5%–10 %) at 0 eV. The static values of energy loss with all applied strain were observed zero. The energy loss starting points in the visible region and got the highest values from 0 eV to 15 eV as shown in **Figure 6**. The extinction coefficient values starting from 0.55(0% strain) with 8.12eV and the highest value goes till 1.33 (+5%) with 7.49eV. The static values of reflectivity with all applied strain were observed as zero. The reflectivity starting points in the visible region got the highest values from 7.01eV-8.71eV as shown in the **Figure 6**. Furthermore, static values of refractive index of strained MgO monolayer are observed to be 1.44(-10%), 1.46(-5%), 1.47(-2%), 1.24(0%), 1.49(+2%), 1.52(+5%), 1.56(+10%).

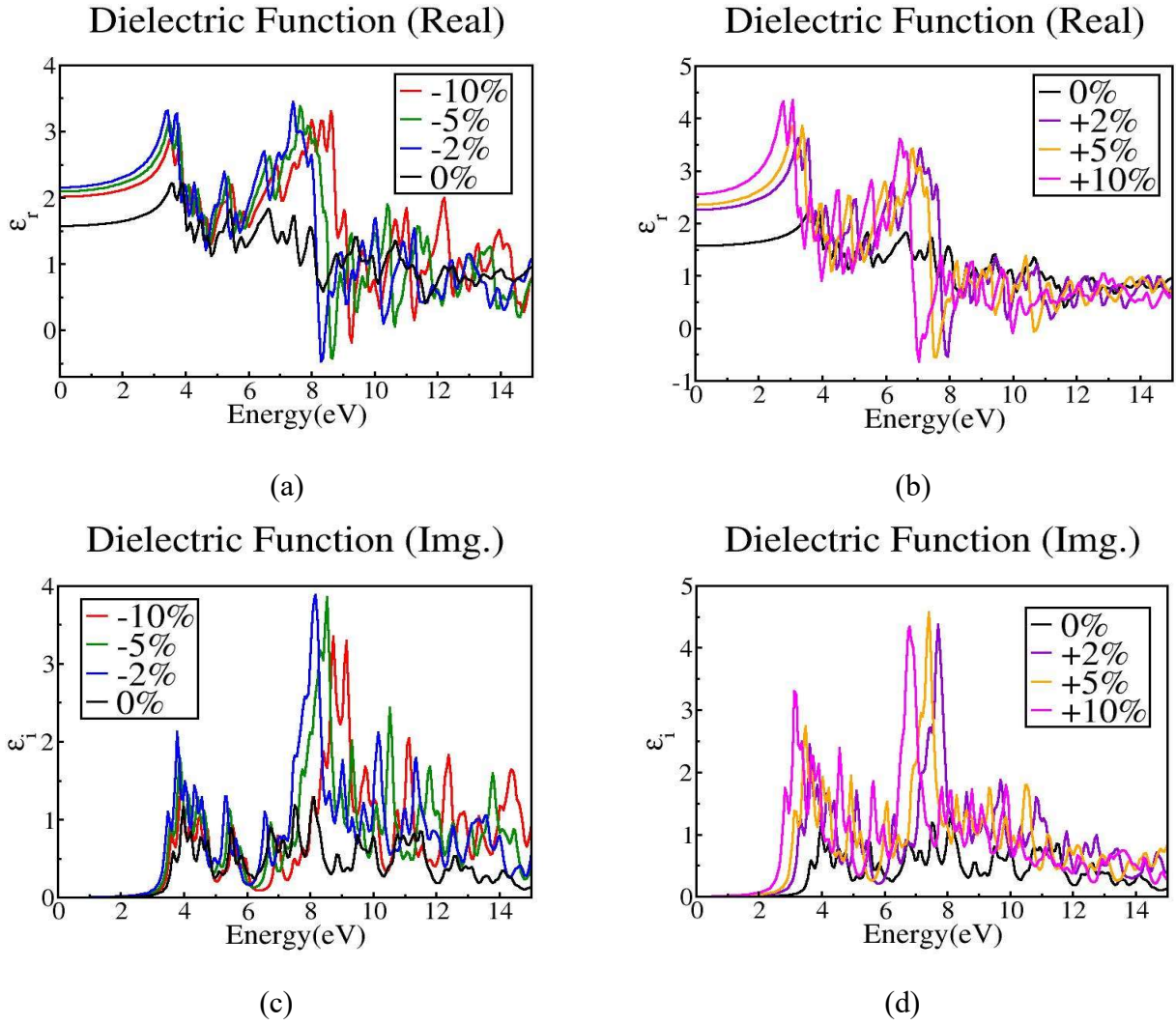

**Figure 4** Real and imaginary components of complex dielectric functions at different biaxial strain values

**Table 2** Optical and dielectric properties of MgO monolayer

| Strain | α(ω) | L(ω) | n(ω) | R(ω) | k(ω) |
|---|---|---|---|---|---|
| **-10%** | 0.000146091 | 9.56176E-07 | 1.44609 | 0.033299523 | 1.4E-06 |
| **-5%** | 0.000176715 | 1.11472E-06 | 1.46307 | 0.03535982 | 1.7E-06 |
| **-2%** | 0.000205599 | 1.26008E-06 | 1.47682 | 0.037064567 | 2E-06 |
| **0%** | 0.000122908 | 1.25412E-06 | 1.24614 | 0.012018413 | 1.2E-06 |
| **+2%** | 0.000261483 | 1.52918E-06 | 1.49997 | 0.039997004 | 2.6E-06 |
| **+5%** | 0.000322096 | 1.80594E-06 | 1.52153 | 0.042787587 | 3.2E-06 |
| **+10%** | 0.000472748 | 2.44169E-06 | 1.56557 | 0.04864249 | 4.7E-06 |

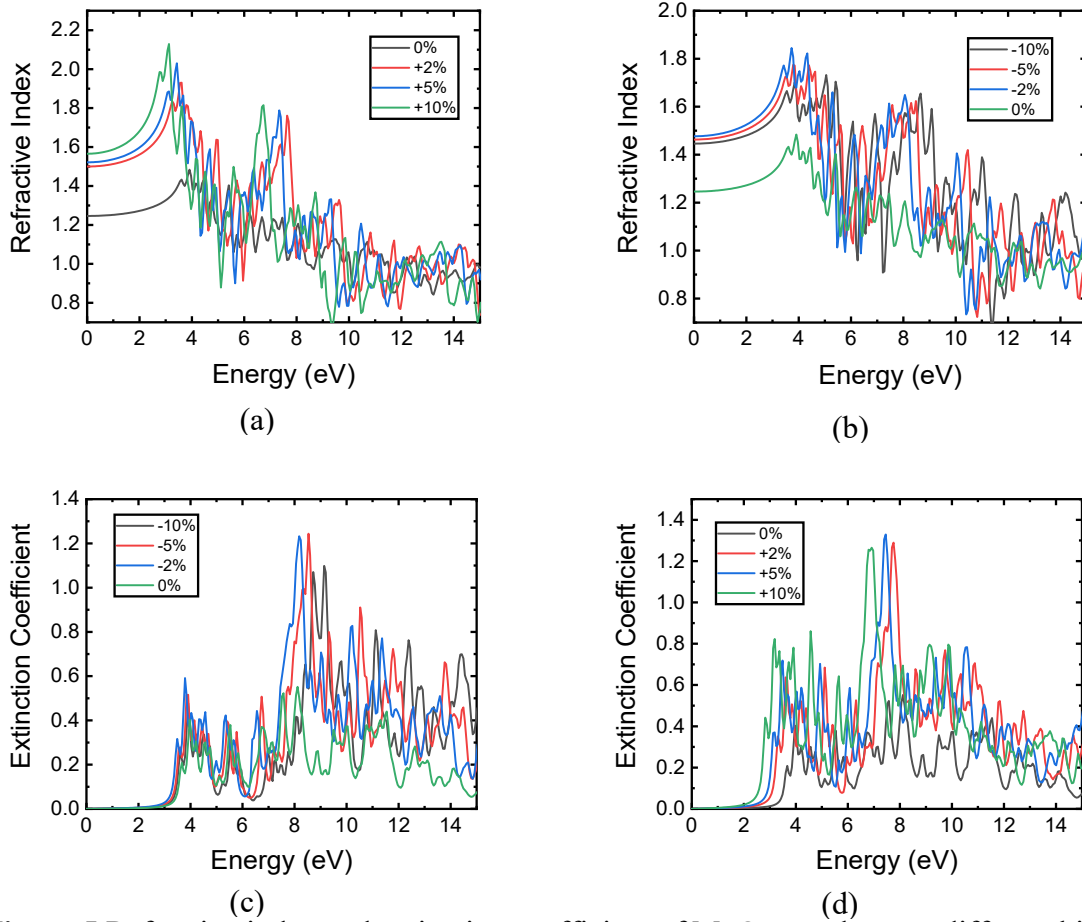

**Figure 5** Refractive index and extinction coefficient of MgO monolayer at different biaxial strain values

**Table 3** Position of highest peak and corresponding energy values in different optical parameters

| Strain | $\varepsilon_r$ | $\varepsilon_i$ | $n(\omega)$ | $\alpha(\omega)$ ($10^6$) | $k(\omega)$ | $L(\omega)$ | $R(\omega)$ |
|---|---|---|---|---|---|---|---|
| **-10%** | 3.30 (8.66eV) | 3.34 (8.81eV) | 1.73 (5.06eV) | 1.02 (9.14eV) | 1.10 (9.14eV) | 2.14 (7.23eV) | 0.20 (8.71eV) |
| **-5%** | 3.37 (7.67eV) | 3.84 (8.59eV) | 1.77 (3.79eV) | 1.08 (8.54) | 1.24 (8.54eV) | 1.37 (16.16eV) | 0.23 (8.54eV) |
| **-2%** | 3.42 (7.45eV) | 3.91 (8.22eV) | 1.84 (3.74eV) | 1.02 (8.18eV) | 1.23 (8.18eV) | 1.39 (17.86eV) | 0.23 (8.22eV) |
| **0%** | 2.22 (3.91eV) | 1.29 (8.21eV) | 1.48 (3.92eV) | 5.13 (11.62eV) | 0.55 (8.12eV) | 0.82 (11.66eV) | 0.07 (8.18eV) |
| **+2%** | 3.64 (3.29eV) | 4.38 (7.74eV) | 1.92 (3.58eV) | 1.01 (7.73ev) | 1.29 (7.76eV) | 1.78 (5.99eV) | 0.24 (7.86eV) |
| **+5%** | 3.91 (3.15eV) | 4.57 (7.48eV) | 2.03 (3.43eV) | 1.00 (7.46eV) | 1.33 (7.49eV) | 1.91 (5.72eV) | 0.26 (7.46eV) |
| **+10** | 4.36 (2.82eV) | 4.33 (6.86eV) | 2.13 (3.12eV) | 8.80 (6.97eV) | 1.26 (6.92eV) | 1.32 (5.14eV) | 0.24 (7.01eV) |

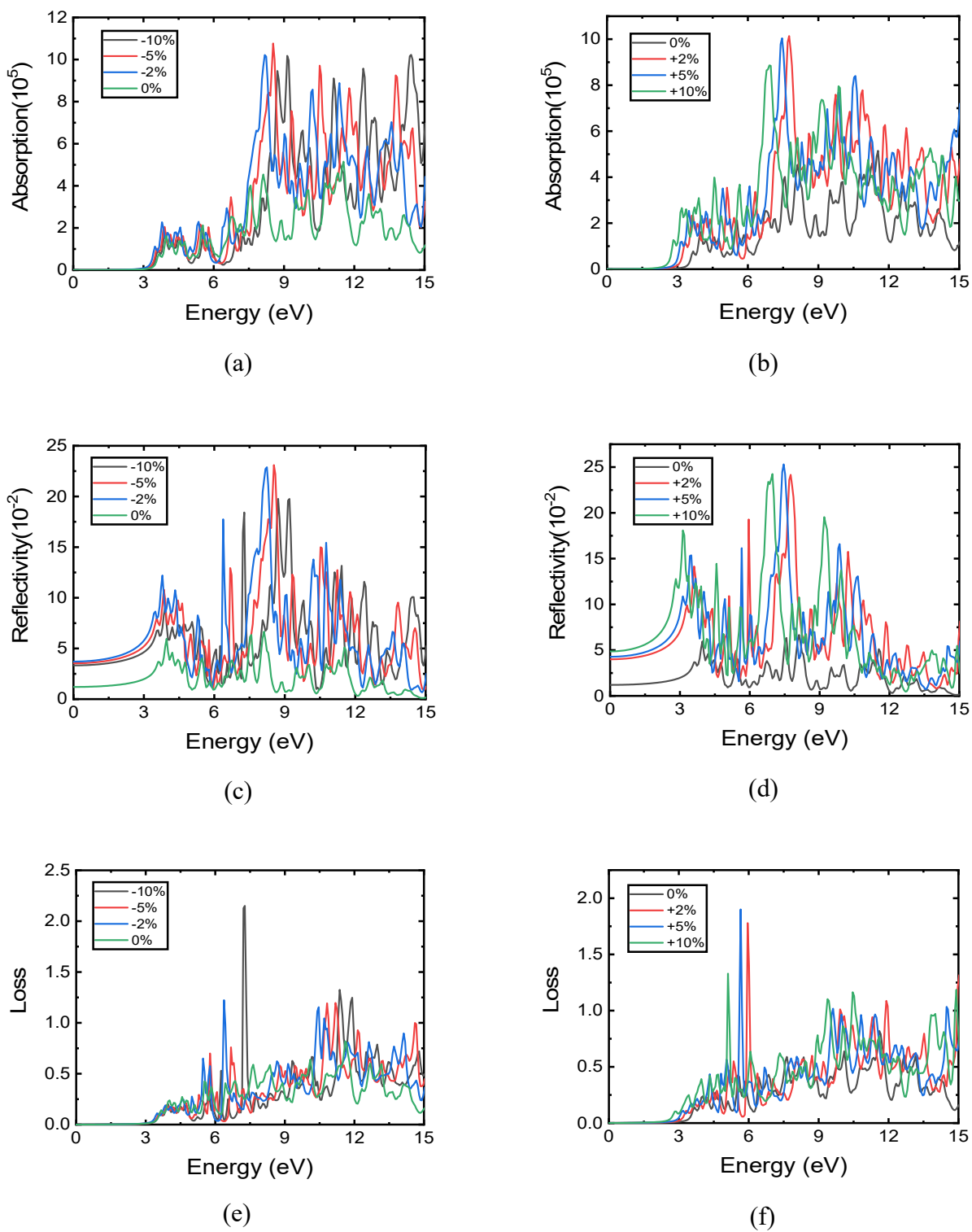

**Figure 6** Average reflectivity, absorption and loss of MgO monolayer at different biaxial strain values

**Conclusion**

In summary, the geometrical structure, distribution of energy levels in the electronic band structure, and optical parameters of 2D MgO monolayer are studied under uniform biaxial (in-plane and out-plane) mechanical strain within a strain interval of -10% to +10% through GGA+PBE method within density functional theory (DFT) framework. The graphene-like hexagonal honeycomb 2D MgO monolayer exhibits an indirect band gap and semiconducting nature. The 2D MgO monolayer maintains its metallic and planar nature even after applying biaxial mechanical strain. Although variations in bond length between Mg and O atoms takes place under strain. To explain the interaction of radiation with 2D MgO monolayer, its optical properties are studied for photon energy between 0 to 15 eV. The position of peaks of different dielectric properties of MgO monolayer and the shift in these peaks induced by biaxial mechanical strain are discussed in detail. Our findings are expected to researchers towards various applications of MgO monolayer in optoelectronics.

**Computational Details**

In our theoretical investigation, we have conducted DFT [79]based calculations through Quantum Espresso (QE) [80]software package. Atomic interactions are described by Generalized Gradient Approximation (GGA) [81]along with Perdew-Burke-Ernzerhof (PBE) exchange-correlation functional[82]. Projector-augmented wave (PAW) pseudopotential is utilized to explain the ion-explain interaction and reduce the computational cost [83]. The unit cell of MgO is composed of one Mg and one O atom (black rectangle in **Figure 1**). A well converged kinetic energy cutoff of about 800 eV is implemented for better accuracy. A 5×5×1 Monkhorst-Pack k-point mesh is used to achieve sampling of Brillouin zone. Also, grimme d3 correction is applied to consider van der Walls interactions and a vacuum of 15Å is implemented to avoid interaction among periodic images. Self-consistence convergence when the force acted on the atoms is reduced to 0.01 eV and energy threshold is set to $10^{-5}$ eV. These computational parameters have helped us in thoroughly analyzing the properties of MgO monolayer and provided important insights to our research

**Acknowledgement**

Dr. Abhishek Kumar Mishra (AKM) expresses sincere gratitude to SERB (Science and Engineering Research Board) for SERB SURE (State University Research Excellence) research grant (SUR/2022/004935). Kamal Kumar (KK) is grateful to UPES (University of Petroleum and Energy Studies) for PhD fellowship.